\documentclass[a4paper,12pt]{article}
\usepackage{graphicx,epsfig,epsf,amssymb}
\usepackage{times}
\newcommand{\degree}{^\circ}
\def \<{\langle}
\def \>{\rangle}

\title {Thermal Gravitational Radiation\\ and Condensed Universe}
\author
{Ti-Pei Li$^{1,2,3}$ and Mei Wu$^2$}
\date{}

\begin{document}
\baselineskip 24pt
\maketitle

{\footnotesize
\begin{enumerate}
\item{Tsinghua Center for Astrophysics, Tsinghua University, Beijing, China
(litp@tsinghua.edu.cn)}
\item{Key Laboratory of Particle Astrophysics, Institute of High Energy Physics,\\ Chinese Academy of Sciences, Beijing, China}
\item{School of Physics, University of Chinese Academy of Sciences, Beijing, China}
\end{enumerate}
\abstract{ The perfect Planck spectrum of the observed cosmic microwave background  radiation indicates that our universe must be in thermal equilibrium. The dark sector of the universe should also be in the same equilibrium state with dark matter and dark energy  coupled to each other and  emits  gravitational  phonon blackbody radiation which is the main component of the cosmic background radiation. In the  radiation-dominated era  such gravitational radiation should be  the majority species of the cosmic medium.   Instead of the ideal fluid assumed by the standard cosmological model $\mathnormal\Lambda$CDM, the universe has to be taken as a thermodynamic system consisting of gravitationally connected dark energy and matter.
Besides particle dynamics,  statistical thermodynamics is also necessary for understanding the cosmological constitution and evolution history.
As an alternative to $\mathnormal\Lambda$CDM we constructed a dark-energy-matter-coupled (DEMC) cosmological model.
Based on the relativistic mass-energy relation, conservation law of energy,  Lagrange's equation with variable potential function, mean-field theory of continuous phase transition, and the symmetry principle of the kinetic coefficients, we deduced dynamic equations of the expansion of  a DEMC universe with three parameters.  These equations  reproduce the observed history of the rate of expansion  of our universe.
}\\

\vspace{8mm}

The detection of  electromagnetic waves (EW) from moving charges by Hertz in 1887 is revered as a great discovery demonstrating  that "electricity has thus annexed the entire territory of light and radiant heat"$^{\scriptsize\cite{caj33}}$.  Thermal radiation emitted from objects with temperature has been  seen to be thermal electromagnetic waves (thermal EW)  because only  EW can travel through empty space. Recently, the event GW150914 observed by LIGO detectors$^{\scriptsize\cite{abb16}}$  shows that, besides EW, gravitational waves (GW) also take place in vacuum spacetime.  A  question naturally arises: whether there exists thermal gravitational radiation (thermal GW)?

We explore here a positive answer. The ordinary matter  constitutes only a small part of total mass in the universe,  the majority of which is in the dark sector, including dark matter and dark energy, which are not coupled to photons.   As shown by the blackbody spectrum of  the cosmic microwave background radiation (CMB) measured by $COBE$, $WMAP$ and $Planck$ satellites, the matter in our universe, including ordinary matter and dark sector,  should be locked  together in thermal equilibrium  with the same temperature of the blackbody  cosmic background radiation (CBR).  However,  only a small fraction of CBR can be attributed to thermal EW, because  most of the matter in the universe cannot  get  involved  in   electromagnetic processes at all. Most CBR should be coupled with the dark sector, which can only be involved in thermal processes through gravitational interaction.   Thus a natural  candidate for  the main component of CBR  should be the thermal GW coupled with dark matter and dark energy through gravitational interaction.\\

\noindent{\large\bf Cosmic  Gravitational Radiation}\\
\noindent{\bf Thermal gravitational phonons}.  One may suggest  that the dark sector could not emit any radiation and CBR is composed of only thermal photons from ordinary matter. Such an assumption leads to grave inconsistencies in  cosmology and fundamental physics: The radius $R$ of the universe increases with the cosmic expansion, the energy density of non-relativistic matter falls as $\rho_m\propto R^{-3}$, while the temperature of   CBR  decreases as $T\propto R^{-1}$,   the energy density of CBR falls as $\rho_{rad}\propto R^{-4}$, then the energy density ratio of radiation  to matter decreases with expansion as $\rho_{rad}/\rho_m\propto R^{-1}$, thus the early universe was dominated by thermal radiation$^{\scriptsize\cite{pee93}}$.   For example,   at the time $t\sim 1$\,s after inflation, temperature $T\sim 10^{10}$\,K ($kT\sim 1$\,MeV), the ratio $\rho_{rad}/\rho_m$ estimated by the standard model of cosmology $\mathnormal\Lambda$CDM  is as  high as $\simeq 6\times 10^5$,  and the universe consisted of almost all thermal radiation.  If the thermal radiation  in early universe was composed of only photons,  almost no mater could exist at the present  unless the thermal radiation at $t\sim 1$\,s consists of relativistic particles with mass much less than 1\,MeV,  which is conflict with particle  experiments.

   To explain why the dark sector can produce thermal gravitational radiation,  we examine electromagnetic waves. The EW detected by Hertz is a solution of  the Maxwell electromagnetic field equations in free space, which,  being a kind of non-thermal radiation, can be produced by a single positive or negative charge moving in acceleration.  On the other hand, thermal energy contained in an electrically neutral object of condensed matter composed of unit cells with interacting positive and negative charged particles in equilibrium, represents vibrational motion of  the elastic structures. Phonons -- quasi-particles as collective excitations of vibrations of elastic structures -- play a major role in heat conduction in an object. Thermal phonons can be generated from different  processes.  For example, optical phonons are generated from vibrational modes  involving oscillating electric dipoles while the center of mass of a unit cell remains static and the atoms within the cell vibrate against each other. Such phonons  can be coupled directly with  electromagnetic fields.  An intrinsically different kind of phonons, the so called "acoustic phonons",  are elastic waves from vibrational  modes  with all atoms in the unit cell vibrating in phase and with the same amplitude$^{\scriptsize\cite{gro14}}$.  Basic thermodynamic properties in such an object are irrelevant to the particular nature, optical or acoustic, of phonons.  Photons and different kinds of phonons  can be coupled to each other in a condensed matter object, but it is not quite clear how phonons (mechanical waves) pass through the boundary of the object into empty space  and completely transform to photons  (electromagnetic waves).

 The recent detected GW is predicted by Einstein's gravitational field equations for accelerated moving masses, that should be a kind of non-thermal radiation.
 It has been thought that gravity cannot be thermalized$^{\scriptsize\cite{pen04}}$, because only positive gravitational mass and only attractive interaction between masses exist in gravity, while in EM exist both positive and negative charges, and both attractive force (between opposite charges) and repulsive force (between  the same  charges). However, there exists dark energy with repulsive gravitation as another basic component of the universe  besides matter with attractive gravitation. The dark energy can be seen as a kind of matter with negative gravitational mass, and  there are also both attractive gravitational force (between dark matters) and repulsive gravitational force (between dark matter and dark energy or between dark energies) in cosmological scales.  The dark sector of the universe is  constituted by two coupled fields, in which elastic structures  at the microscopic scale can be formed from quantum fluctuations. The CMB radiation with most perfect blackbody spectrum ever observed in nature further demonstrates  that the universe is in full thermal equilibrium,
which requires that gravity must dominate the cosmological thermalization, particularly in the early universe.
The two kinds of gravity, attractive force energized by matter and repulsive force by dark energy, like that of  electromagnetic interaction by two opposite  charges, can closely coupled to keep the universe in thermal equilibrium.
   Analogous to an electrically neutral object producing thermal phonons  inside and radiating out  thermal photons, the dark sector of the universe, being  gravitationally neutral and thermally equilibrated, would produce thermal phonons,  within which we, the observers, are immersed.  Therefore,   cosmic gravitational phonons as the global thermal carriers  of  the universe should  coexist in CBR  with thermal photons  produced and traveled  locally in  bounded  systems of  baryonic matter.

We will show in the next section that the global energy conservation can be recovered for the whole universe by including gravitational phonons.\\

\noindent{\bf Global conservation of energy}.  The density of matter  of a homogeneous and isotropic universe  is the global mean of all types of energies in all local systems, e.g.  mass in galaxies, kinetic energy and pressure of stars and galaxies,  intergalactic gas, cosmic rays, magnetic fields, radiation etc (\cite{mis73}, \S27.2).  In  the coordinates comoving with an expanding  universe,  radiation energy is part of  rest energy.  The total rest energy  in natural units ($c=1$) can be expressed as
  \begin{equation}\label{eq:Erest}
  E_{rest}=M+E_{rad}\,,
  \end{equation}
   where  the total radiation energy $E_{rad}$ decreases as $1/R$ as the universe expands,  the total rest mass
   \begin{equation}\label{eq:M}
   M=M_{m}+M_{\lambda}  \end{equation}
    with $M_m$ and $M_\lambda$ being rest mass of matter (dark matter and ordinary matter) and of dark energy, respectively.

     The  global conservation of energy is a difficult issue in relativistic cosmology.  From $t\sim 4\times 10^5$\,y, when $E_{rad}/M\sim 10^{-4}$ and photons decoupled  from matter and started to travel freely in the universe,   no more radiation energy can  be converted into matter through photon-matter interaction. Where does the lost radiation energy go?  Even more puzzling than the loss of  $10^{-4}E_{rest}$ is the disappearance of almost all energy during expansion.  At the early time $t\sim$ 1\,s, $E_{rad}/M\sim 6\times 10^{5}$, while matter can be ignored,  the photon energy already decreased down to $kT\sim 1$\,MeV.  Since that time, due to the inability of photons' converting into particles through pair production, radiation energy should be lost along with the expansion of the universe.

     The global energy paradox in cosmology can be naturally solved with considering the gravitational phonons.  The radiation energy of the dark sector is a collective energy since cosmic gravitational phonons represent mechanical oscillations  of the elastic structures throughout the dark sector.  Gravitational phonons  connected with dark matter and dark energy cannot be taken as an isolated system.
     In addition, the successful detection of GW by means of optical  interferometer demonstrates that laser photons propagating in vacuum can be effectively  modulated by GW, or, in other words,  there exists  interaction between photons and  gravitational radiation in vacuum$^{\scriptsize\cite{ger62}}$.
     During the cooling of the expanding universe,   energy of photons even after decoupling from ordinary matter could be  transformed into gravitational phonons through photon-phonon interaction, and  gravitational radiation energy  could be continuously  transformed into the rest mass of the elastic system through phonon-matter connection, thus conserving the total rest energy
 \begin{equation}\label{eq:Erest}
 E_{rest}=\mbox{const}\,, \end{equation}
  as schematically shown in Figure \ref{Fig:Erest}. The  density of rest energy (inertial density) can be written as
  \begin{equation}\label{eq:ri}
    \rho=\rho_{_m}+\rho_{_\lambda}+\rho_{_{rad}}=\rho_{_0} a^{-3}\,,
    \end{equation}
  where $\rho_{_m}$ and $\rho_{_\lambda}$ are the rest mass density of matter and of dark energy, respectively;  $\rho_{_{rad}}$ is the density of radiation energy;  $\rho_{_0}$  the rest energy density of the current universe, and the scalar factor $a=R/R_{_0}$ with $R$ and $R_{_0}$  being the radius of the universe and the current radius, respectively.\\
 \begin{figure}[t]
   \begin{center}
   \includegraphics[width=100mm,height=80mm,scale=1.0,angle=0]{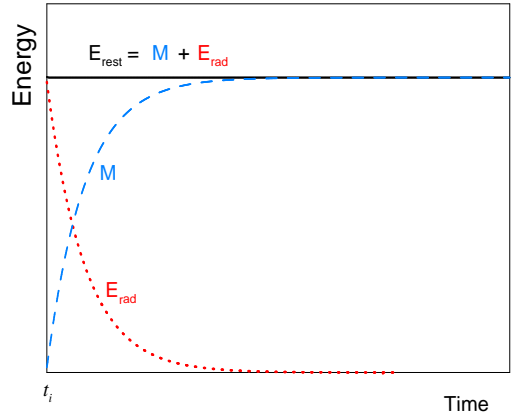}
   \vspace{-1.6cm}
   \caption{ \label{Fig:Erest} Evolution of  total rest energy $E_{rest}$, rest mass of matter $M$, and radiation energy $E_{rad}$ of expanding universe after the end time $t_i$ of inflation.}
   \end{center}
\end{figure}

In the lecture  $\ll$Statistcal Thermodynamics$\gg$, Schr\"{o}dinger pointed out:   In Boltzmann-Gibbs Statistics,  an ideal gas system  at ordinary temperature would entail extreme rarefication in the case of allowing  particles to be created or annihilated in collisions,  or to transit  into heat radiation.  "Unless we want to assume that there is no much ponderable matter left in the universe as there is. The only way out seems to be to assume that the transition is a very slow process and that not very far back the condition of the universe were very different from what they are now."$^{\scriptsize\cite{sch48}}$  Recently,  it is realized in thermal engineering that Fourier¡¯s law of heat conduction no longer holds when the heat flux is very high.  Guo et al.$^{\scriptsize\cite{guo07,cao07}}$ pointed out that the failure of  Fourier¡¯s law comes from ignoring the inertial force of the phonon gas, and established an improved equation of motion for the phonon gas in a solid including the equivalent mass of thermal vibration energy of lattices. They submit  a question to physics: where the thermal energy of the equivalent mass of the phonon gas go during heat dissipation?
It seems that, to maintain the law of conservation of energy, the energy transformation from phonons back to matter is needed  not only in cosmology, but also in thermodynamics.

In 1920s, in microscopic physics there was a puzzle of  non-conservation of energy in $\beta$-decay. Two different explanations about the puzzle were proposed:  Bohr  claimed   that energy was not  necessarily conserved in quantum theory, while Pauli in 1930 suggested that a new type of particle (neutrino) accounting for the missing energy was also emitted during beta decay, which was later discovered. Now in cosmology there exists  an analogous puzzle of non-conservation of radiation energy in the expanding universe.  The standard model of cosmology has circumvented this problem by requiring no global energy conservation  in curved  spacetimes of general relativity theory (GR)$^{\scriptsize\cite{pee93}}$.  However,  the space of our universe is actually very flat.   CMB and BAO (Baryon Acoustic Oscillations) measurements$^{\scriptsize\cite{pla13}}$ give a very  small  spatial curvature for the present universe $\Omega_{k}=0.000\pm 0.007$, which constrains early universe to be even more strictly  flat, e.g.   $\Omega_{k}<10^{-58}$ during the Planck era. In addition, a given moment of time can be uniquely determined  by a  given hypersurface in the comoving, synchronous coordinate system established for a homogeneous and  isotropic universe (\cite{mis73}, \S27.3). Combined with the fact that  the transitivity of thermal equilibrium is equivalent to the transitivity of clock rate synchronization$^{\scriptsize\cite{zao99}}$, a unique time can be defined for the universe.
 The flatness of our universe as a Euclidean spacetime has been validated by observations. It necessarily follows that
 the global energy of the universe must be conserved and there is no reason to break the conservation law in cosmology through the means of curved spacetime or GR.  The puzzle of  non-conservation of energy in cosmology can be solved by incorporating the thus-far ignored matter species, namely, gravitational phonons,  which are in thermal equilibrium with coupled matter and dark energy in the flat spacetime of the universe.\\

\noindent{\large\bf Cosmic Expansion Dynamics}

\noindent The extremely perfect Planck spectrum of CMB strictly constrains the content and form of the cosmic medium.  Blackbody radiation can be constructed by two compositions in thermal equilibrium:  bosons  (photons and/or phonons) with zero rest mass as hot carriers, and condensed matter providing microscopic oscillators.  CBR's  accurately following the Planck's law of blackbody radiation requires  cosmic media  be  like a condensed matter consisting of elastic oscillators with balanced attractive and repulsive gravity. Therefore, the universe is a collective system, the cosmic dynamics should be based on the equilibrium thermodynamics.
In \S22.2 of the textbook $\ll$Gravitation$\gg$ by Misner, Thorne \& Wheeler, after discussing the equilibrium thermodynamics of a perfect fluid with thermodynamic variables $\rho$ (total mass-energy containing in a unit volume in the rest frame of the fluid), $\mu$ (chemical potential in rest frame)  etc,  the authors  concluded: "All the above laws and equations of thermodynamics are the same in curved spacetime as in flat spacetime, and the same in (relativistic) flat spacetime as in classical non-relativistic thermodynamics -- except for the inclusion of rest mass together with all other forms of mass-energy, in $\rho$ and $\mu$. The reason is simple: the laws are all formulated as scalar equations linking thermodynamic variables that one measures in the rest frame of the fluid."$^{\scriptsize\cite{mis73}}$ Accordingly, we now derive the dynamic equations of cosmological expansion in classical Galilean spacetime with the additional requirement that all formes of mass-energy are included into rest mass.\\

\noindent{\bf Expansion equations}.  For a homogeneous and isotropic universe, one can define a comoving coordinate system with the Robertson-Walker metric. Cosmic dynamics studies the expansion of the comoving system itself. Matter in cosmic dynamics, as indicators of the cosmic expansion, have to keep rest with respect to the comoving coordinate system, its rest energy should include all forms of mass-energy.
  We can take the comoving system at the moment when inflation initiated and with respect to which the expansion rate $\dot{R}$ of the universe is measured. In the initial static system, the kinetic energy of the expansion can be expressed as
\begin{equation}\label{eq:Ek}
E_k=E_{rest}\,\dot{R}^2/2\,, \end{equation}
where  the total inertial mass $E_{rest}$ measured in the comoving coordinates of the universe includes all forms of mass-energy.

For a body with mass $M$ and radius $R$, the gravitational potential function is
\begin{equation}\label{eq:ep0}
\mathnormal\Phi=\frac{GM}{R}\,, \end{equation}
where $M$  is not the inertial mass but gravitational mass, although they are identical  to each other in both Newtonian and GR frameworks. However, to be a source of gravity  the gravitational mass of repulsive dark energy should be negative, in consequence the gravitational mass $M_g$ of the universe should be the net of matter and dark energy, i.e.
\begin{equation}\label{eq:Mg}
M_g=M_m-M_{\lambda}\,. \end{equation}
Accordingly,  the gravitational mass density
    \begin{equation}\label{eq:rg}
    \rho_{_g}=\rho_{_m}-\rho_{_\lambda}\,. \end{equation}
The gravitational potential function of the universe should be expressed as
 \begin{equation}\label{eq:phi}
 \mathnormal\Phi=\frac{GM_g}{R}\,. \end{equation}
The potential energy of the universe  then becomes
\begin{equation}\label{eq:Ep}
E_p= -E_{rest}\frac{GM_g}{R}\,. \end{equation}

 Although the above expressions of kinetic and potential energies of expanding universe, Eq.\,(\ref{eq:Ek}) and Eq.\,(\ref{eq:Ep}),  are  formulated  in Galilean coordinates,  they cannot be simply thought of as Newtonian ones, because  the rest masses $E_{rest}$, $M_m$ and $M_\lambda$    in these formulae include all forms of mass-energy in local Minkowski  spacetime,  and are not classical quantities in Newtonian framework. Furthermore, the definition  of gravitational mass (Eq.\,\ref{eq:Mg}) is also not a Newtonian one.

 From the conservation of mechanical energy
 \begin{equation}\label{eq:mech}
  E_{mech}=E_k+E_p=\mbox{const}\,, \end{equation}
  we get the energy equation of the expanding universe
\begin{equation}\label{eq:ee}
\dot{R}^2= 2G M_g R^{-1}+E^\prime  \end{equation}
with the constant $E^\prime=2E_{mech}/E_{rest}$, or equivalently
 \[ \hspace{5.8cm}
\dot{a}^2=\frac{8\pi G}{3} \rho_{_g}  a^2+\epsilon\,, \hspace{5.3cm} (\ref{eq:ee})^\prime \]
with  the constant
\begin{equation} \epsilon=\frac{2E_{mech}}{R^{^2}_{_0}E_{rest}}\,. \end{equation}

For gravitational field of massive matter,  the zero point of the potential function Eq.\,(\ref{eq:ep0}) is artificially chosen at $R=\infty$ and the potential energy  is then  always less than zero, which has been used by some cosmologists to claim that the universe is the ultimate "free lunch",  provided the gravitational  potential energy is negative  and the total energy of the universe is zero$^{\scriptsize\cite{haw05}}$.
However,  there is no such an arbitrariness of choosing potential zero point and thus no "free lunch" for a coupled field of matter and dark energy: zero potential $\mathnormal\Phi=0$ in Eq.\,(\ref{eq:phi}) is physically determined  by matter-dark energy equilibrium at $M_g=0$, i.e. $M_m=M_\lambda$,  and negative or positive potential energy are corresponding to attraction or repulsion domination, respectively.  Increasing (or decreasing) of $M_g$  will cause $\dot{R}$ to decrease (or increase) for maintaining energy conservation in  Eq.\,({\ref{eq:ee}).

With Eqs.\,(\ref{eq:Ek}) and (\ref{eq:Ep}) the Lagrangian can be written out as
\begin{equation}\label{eq:L}
{\cal L}=E_{rest} \dot{R}^2/2+E_{rest}G M_g  R^{-1}\,. \end{equation}
The equation of motion of an expanding universe can be derived by Lagrange's equation
 \begin{equation}\label{eq:le}
 \frac{\partial {\cal L}}{\partial R}-\frac{\mbox{d}}{\mbox{d}t}(\frac{\partial {\cal L}}{\partial \dot{R}})=0\,.
 \end{equation}
     For a universe with fixed composition,  i.e. $M_g=\mbox{const}$,  Eq.\,(\ref{eq:le}) reads
\begin{equation}\label{eq:me0}
\ddot{R}=- G  M_{g} R^{-2}\,, \end{equation}
or
\[ \hspace{6.3cm} \ddot{a}=- \frac{4\pi G}{3} \rho_{_g} a\,. \hspace{5.4cm} (\ref{eq:me0})^\prime \]

From Eqs.\,(\ref{eq:ee})$^\prime$ and (\ref{eq:me0})$^\prime$ we see that a matter-dark energy balanced universe with $\rho_{_g}=0$ will experience a uniform expansion with a constant rate $\dot{a}=\sqrt{\epsilon}$
and $\ddot{a}=0$, and a dark energy (or matter) dominated  universe of $\rho_g<0$ with $\dot{a}<\sqrt{\epsilon}$ (or $\rho_g>0$ with $\dot{a}>\sqrt{\epsilon}$), the expansion rate is
ever increasing  (or decreasing)  toward an asymptotic value of  $\sqrt{\epsilon}$.

  An evolution equation for a universe with fixed composition can be derived by combining Eqs.\,(\ref{eq:ee}) and (\ref{eq:me0}) as follows
\begin{equation}\label{eq:ev0}
\ddot{R}=-\frac{1}{2}(\dot{R}^2-E^\prime)\,R^{-1}\,, \end{equation}
or in scalar factor
\[ \hspace{5.8cm}
\ddot{a}=-\frac{1}{2}(\dot{a}^2-\epsilon)\,a^{-1}\,.  \hspace{5.2cm} (\ref{eq:ev0})^\prime \]
The  evolution equation  (\ref{eq:ev0})$^\prime$ is a type of Bernoulli's equations in fluid dynamics with three solutions:
\begin{eqnarray}
\dot{a} & =& \sqrt{c_{_1}(1+z)-\epsilon} \hspace{5mm} \mbox{for}~\dot{a}>\sqrt{\epsilon} \hspace{6mm} (\ddot{a}< 0,~ \mbox{deceleration}) \label{s:dec0}\\
\dot{a} & =& \sqrt{\epsilon} \hspace{25mm} \mbox{for}~\dot{a}=\sqrt{\epsilon} \hspace{6mm} (\ddot{a}=0, ~   \mbox{constant expansion}) \label{s:con0} \\
\dot{a} & =& \sqrt{\epsilon-c_{_2}(1+z)} \hspace{5mm} \mbox{for}~\dot{a}<\sqrt{\epsilon} \hspace{6mm} (\ddot{a}> 0,~ \mbox{acceleration}) \label{s:acc0} \end{eqnarray}
where $c_{_1}$ and $c_{_2}$ are positive integral constants, and $a=(1+z)^{-1}$.
  Eqs.\,(\ref{s:dec0})-(\ref{s:acc0}) show again that the expansion rate $\dot{a}$ of an accelerating (or decelerating)  universe without phase transition will be monotonically increasing (or decreasing) to approach to $\sqrt{\epsilon}$ with $a\rightarrow \infty$ ($z\rightarrow -1$). \\

\noindent{\bf Expansion with phase transition}.
  In the end of 1990's,  the observations of  Type Ia supernovae$^{\scriptsize\cite{rie98,per99}}$ found that the current cosmic expansion is to be  accelerating, and later observations$^{\scriptsize\cite{rie01,rie04}}$  gave the evidence for  a decelerating phase being  proceeded the accelerating epoch at $z\sim 0.5$, which  indicate that our universe seems to have
 undergone a phase transition.  Since then, based on CMB maps, BAO features,  and the galaxy differential age method,  Hubble parameters $H(z)$, or,  expansion rates $\dot{a}(z)=H(z)/(1+z)$,   have been measured  for different redshift $z$, the current available observational results are shown in Figure\,\ref{fig:ad-z}.  We can see from Figure\,\ref{fig:ad-z} that the universe  in earlier epoch of $z>z_c\sim 1$  was more likely uniformly expansion.
When $z\le z_c$, or, accordingly $R\ge R_c=R_{_0}/(1+z_c)$ and $T\le T_c=T_{_0}(1+z_c)$,
the universe seems to enter into a period of phase transition: a decelerating phase starting from  $z_{_c}\sim 1$ being succeeded by an acceleration epoch from  $z_{_t}\sim 0.6$.
 During the phase transition period, the composition of the cosmic medium converts  between matter and dark energy ($M_m$ and $M_\lambda$). The changing of the gravitational mass $M_g$  has to be considered in studying  expansion dynamics.

\begin{figure}[t]
   \begin{center}
   \includegraphics[width=135mm,height=110mm,scale=1.5,angle=0]{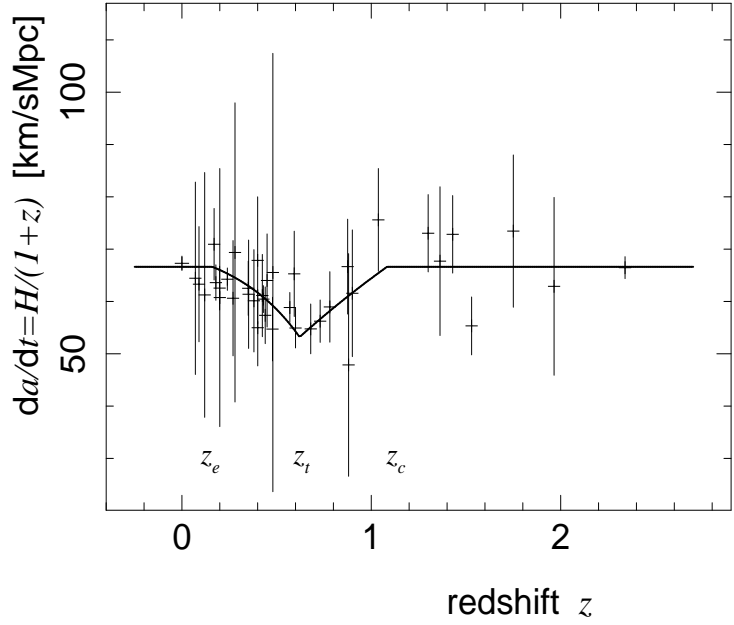}
   \vspace{-1.6cm}
   \caption{ \label{fig:ad-z}  Expansion rate verses redshift.    The Hubble parameters $H(z)$ for $z=0$ is reported by  Planck Collaboration \cite{pla13}, for $0<z\le 2.34$  by \cite{jim03}-\cite{moresco16} and  collected by \cite{dua16}. The line is the fitted result based on Eq.\,(\ref{eq:sp}) of the condensed universe model DEMC. }
   \end{center}
\end{figure}

 The decelerating (decreasing of the kinetic energy)  starting from $z_c$ should be caused by the increasing of the potential energy (decreasing of $M_g$, i.e. partial $M_m$ being translated into $M_\lambda$).  Adopting the methodology of  Landau mean-field theory for continuous phase transition$^{\scriptsize\cite{lan07}}$ we expand  $M_g$  in power series of $(R-R_c)/R_c$ about the transition point $R_c$:
 \begin{equation} M_g=b_{_0}+b_{_1}(R-R_c) +\cdots\end{equation}
  with $b_{_0}=M_g(R=R_c)=0$.  For the decelerating phase of $z<z_c$, due to the rarity and large uncertainty of data, only the first two terms in the power series can be retained
 \begin{equation}\label{eq:b}
 M_g=-b (R-R_c) \end{equation}
 with $b>0$.
From  Lagrange's equation (\ref{eq:le}) and $\partial M_g/\partial R=-b$
we get the equation of motion of the expanding universe for the decelerating phase  of  $z\le z_c$
\begin{equation}\label{eq:me}
\ddot{R}=- G (M_g R^{-2}+bR^{-1})\,. \end{equation}
The following evolution equation can de derived by combining Eqs.\,(\ref{eq:ee}) and (\ref{eq:me})
\begin{equation}\label{eq:ev1}
\ddot{R}=-\frac{1}{2}(\dot{R}^2-E^{\prime}+2Gb)\,, \end{equation}
or in scalar factor
\[ \hspace{5.7cm}
\ddot{a}=-\frac{1}{2}(\dot{a}^2-\alpha \epsilon)\,a^{-1} \hspace{5.3cm} (\ref{eq:ev1})^\prime \]
with
\begin{equation} \alpha=1-2Gb/R_0^2\,. \end{equation}
Eq.\,(\ref{eq:ev1})$^\prime$ has a  solution
\begin{equation}\label{eq:s1}
\dot{a} = \sqrt{c(1+z)-\alpha \epsilon} \hspace{5mm} ( z_t\le z<z_c)\,, \end{equation}
where  $z_{_t}$  is the end redshift of the decelerating phase, $c$  is a positive integral constant which can be determined from the above equation and $\dot{a}(z=z_c)=\sqrt{\epsilon}$ as follows
\begin{equation}\label{eq:c}  c=\epsilon(1+\alpha)/(1+z_c)\,. \end{equation}

\begin{figure}[t]
   \begin{center}
   \includegraphics[width=95mm,height=80mm,scale=0.3,angle=0]{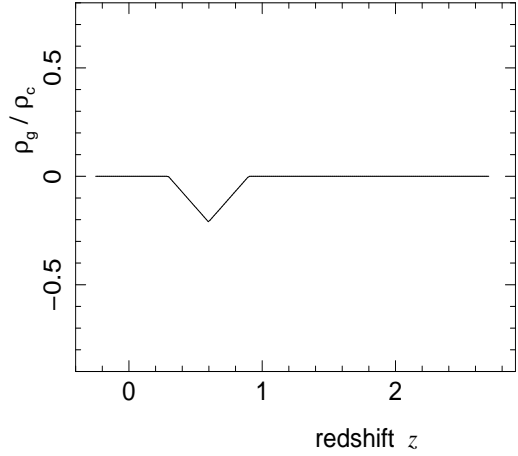}
   \vspace{-1.6cm}
   \caption{ \label{fig:rho} Evolution of  fractional gravitation density $\eta=\rho_{_g}/\rho_{_c}$.}
   \end{center}
\end{figure}

 After the decelerating  phase with matter converted into dark energy, the universe at $z<z_{t}$   relaxed to an equilibrium state again  through an accelerating phase of $\ddot{a}>0$ with dark energy converting back into matter.  For a continuous phase transition at $z_{t}$,  the acceleration   $\ddot{a}(z=z_{t})=0$,    the expansion rate (from Eq.\,\ref{eq:ev1}$^\prime$)
\begin{equation}\label{eq:adzt}  \dot{a}(z=z_{t})=\sqrt{\alpha \epsilon}\,. \end{equation}
Substituting Eq.\,(\ref{eq:adzt})  into Eq.\,(\ref{eq:s1}) we find
\begin{equation}\label{eq:zb}
z_{t}=\frac{2 \alpha}{1+\alpha}(1+z_c)-1\,. \end{equation}
From $0<z_{t}< z_c$ we get
\begin{equation}\label{eq:alpha} \frac{1}{2z_c+1}< \alpha<1\,.\end{equation}
We  define a  dimensionless gravitation density
 \begin{equation}\label{eq:eta}  \eta=\rho_{_g}/\rho_{_c}=\frac{\dot{a}^2}{\epsilon}-1\,,  \end{equation}
 where
  \[  \rho_{_g}=\frac{3}{8\pi G} (\dot{a}^2-\epsilon) (1+z)^2 \]
  is deduced from Eq.\,(\ref{eq:ee})$^\prime$,  and the critical density
 \[ \rho_{_c}\equiv \frac{3}{8\pi G}\,\epsilon\, (1+z)^{3}\,. \]
   By Onsager's principle  of the symmetry of the kinetic coefficients$^{\scriptsize\cite{lan07}}$,  $\eta$  should be symmetric around $z=z_{t}$, i.e. $\eta$ at $z_e\le z<z_{t}$ is the same with  that at
 $ z^\prime=2z_{t}-z$, where
\begin{equation}\label{eq:zcp}  z_e=2z_{t}-z_c\,. \end{equation}
Then  from Eq.\,(\ref{eq:s1})  we obtain
 \begin{equation}\label{eq:s2}
 \dot{a} = \sqrt{c(1+2z_{t}-z)-\alpha \epsilon} \hspace{5mm} (z_e \le z<z_{t})\,. \end{equation}

Therefore, in the condensed universe model, or  dark energy-matter-coupling model (DEMC), the observed  expansion history of our universe can be described by
\begin{eqnarray}
\dot{a} & =& \sqrt{\epsilon} \hspace{43.5mm} z\ge z_c \hspace{5mm} (\ddot{a}=0)\,; \nonumber \\
 &  & \sqrt{c(1+z)-\alpha \epsilon}  \hspace{14mm} z_{t}\le z<z_c \hspace{5mm} (\ddot{a}< 0)\,; \nonumber\\
 &  & \sqrt{c(1+2z_{t}-z)-\alpha \epsilon}\hspace{4mm} z_e\le z<z_{t}\hspace{5mm} (\ddot{a}> 0)\,; \nonumber\\
 &  & \sqrt{\epsilon} \hspace{43.5mm} z< z_e \hspace{5mm} (\ddot{a}=0)\,. \label{eq:sp}
 \end{eqnarray}

 We now fit the DEMC model described by Eq.\,(\ref{eq:sp})  to  the measured data of $\dot{a}(z)$ with the three  parameters $(\epsilon,  \alpha,  z_c)$  and  Eqs.\,(\ref{eq:c}) -- (\ref{eq:zcp}),  the solid line in Figure\,\ref{fig:ad-z} represents  the  result with the estimated values $\sqrt{\hat{\epsilon}}=66.7\pm 0.9$ km s$^{-1}$ Mpc$^{-1}$,  $\hat{\alpha}=0.64\pm 0.04$,  and  $\hat{z}_c=1.08\pm0.15$ ,  where the standard deviations in the above estimations  were derived from bootstrapped data samples produced by Gaussian sampling from the measured data set. The corresponding evolution history of fractional gravitation density $\eta$  is shown in Figure\,\ref{fig:rho}.

 The major difference between DEMC and $\mathnormal\Lambda$CDM is that uniform expansion with equilibrium between matter and dark energy  is ordinary  for former but never occur in later.  In $\mathnormal\Lambda$CDM,  the  universe is always violently  unstable,  the evolution history is presented as an inverted bell-shaped curve as shown by the dashed-dotted curve  in Figure\,\ref{fig:history}:  in the past, the larger the redshift $z$, the higher the expansion rate ($\dot{a}\propto \sqrt{z}$ at $z\gg z_c$ or $a\ll a_c=1/(1+z_c)$), and in the future, the expansion rate will infinitely increase ( $\dot{a}\propto a$ at $a\gg a_c$).  In contrast, a dark energy-matter coupling universe is almost always steady in uniform expansion with a constant rate. More accurate measurements of $\dot{a}(z)$ at high $z$ will unequivocally distinguish DEMC and $\mathnormal\Lambda$CDM.

 Cosmological phase transitions are transformations between the two kinds of gravitational fields with opposite signs.  In contrast with any mechanical exploding, phase transition is capable of producing homogenous expansion without a center for fulfilling  the Cosmological Principle. Moreover, in statistical physics,  for a phase transition with a  critical exponent $\nu>0$,    when $T\rightarrow T_c$ the correlation length$^{\scriptsize\cite{nis11}}$
  \begin{equation}\label{eq:cl}
   \xi\propto |(T-T_c)/T_c|^{-\nu}\rightarrow \infty\,, \end{equation}
  showing that  statistical thermodynamics can provide a proper  framework to investigate the  global non-equilibrium processes in the universe. The cosmological observation can also be a unique approach to study phase transitions and critical phenomena. Figure\,\ref{fig:ad-z}  is based on measured data of expansion rate averaged over the whole sky, comparing the evolution history of  different sky regions, e.g.  the distribution of  $z_c$, will provide a probe to the production and propagation of a cosmological phase transition.\\

\noindent{\large\bf Composition, Origin and Fate of the Universe}\\
\noindent The current contents of the universe derived from CMB temperature maps of both $WMAP$$^{\scriptsize\cite{liu09}}$  and $Planck$$^{\scriptsize\cite{pla13}}$   with Friedmann's  equation in $\mathnormal\Lambda$CDM are:  relative baryon  density $\Omega_b=5\%$, dark matter density $\Omega_{dm}=27\%$, and dark energy density $\Omega_{de}=68\%$.  However,  the fitting process with DEMC model give the turning point of the transition
$ \hat{z}_t=0.61\pm 0.08$ and the end  point $ \hat{z}_e=0.15\pm 0.09$, indicating that
the whole phase transition was already finished, the current universe is again  in a state of uniform expansion with matter and dark energy with $\rho_{g}=0$  or $\rho_{_\lambda}=\rho_{_m}$, as shown in Figure\,\ref{fig:rho}.
 Thus  from  the observed expansion history and DEMC,  the current contents of the universe should be
\begin{equation} \Omega_b+\Omega_{dm}=\Omega_{de}=50\%\,,\end{equation}
which is  completely different from what  given by $\mathnormal\Lambda$CDM.

 \begin{figure}
   \begin{center}
\includegraphics[height=11cm, width=16cm, angle=0]{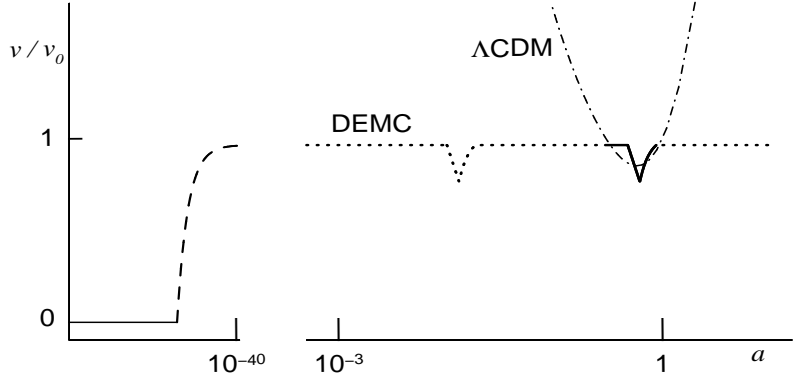}
\vspace{-3.5cm}\caption{\label{fig:history}  A diagrammatic sketch for cosmic evolution. The scalar factor of the universe $a=R/R_{_0}=1/(1+z)$ where $R$ is radius of the universe, $R_{_0}$ the present radius. The expansion velocity $v=\dot{a} R_{_0}$ and $v_{_0}$ the present velocity. The dashed-dotted line sketches for $\mathnormal\Lambda$CDM model derived by fitting observed data of $v/v_{_0}$ between $0.25< a\le1$ $(0\le z< 3)$.
The other  line segments sketch for DEMC model: the thick solid segment is the result of fitting to observed data, dotted segments show prediction of DEMC for earlier and future expansion, dashed and thing solid segments in the very early ages represent inflation and primordially epoches, respectively.
}
\end{center}
\vspace{-7mm}
\end{figure}
Figure\,\ref{fig:history} schematically illustrates the  history of  expansion velocity of a condensed DEMC universe,  where the thick solid line segment represents the already observed part of our universe, dotted line segments are expectations from the scenario of condensed universe for past and future eras -- normally the universe is constantly expanding but  occasionally interrupted by phase transition. We speculatively put a phase transition at $a\sim 10^{-2}$: The large cold spot with angular size $\sim 10\degree$ $^{\scriptsize\cite{cruz07}}$  and quite a few similar cold or hot spots$^{\scriptsize\cite{liu09p}}$   are detected in CMB maps, this may hint  a cosmic perturbation probably having occurred at redshift $z\sim 10^2$.

 The universe in DEMC model was created by an inflation in the primordial vacuum  constituted by universally balanced attractive and repulsive fields possessing only rest energy but no inertia.  The inflation is also a phase transition which broke the equilibrium of a limited region in the static primordial vacuum by converting a significant part of attractive field into repulsive one  and resulted in exponential inflation (the existence of multiple universes originated from different regions of the primordial vacuum is a natural corollary  of  DEMC).  If the inflation occurred in a region of radius $R_{s}=a_{s}R_{_0}$ with a density $\rho_{s}$ of total energy,  from energy conservation, $\rho_{s}=\rho_{_{tot,0}}\, a_{s}^{-3}$, with $0<\rho_{_0}<\rho_{_{tot,0}}$ we get $0<a_{s}<(\rho_{s}/\rho_{_0})^{-1/3}\,.$
 The vacuum energy density $\rho_{s}$ predicted by the uncertainty principle sets an upper limit for the energy density $\rho$.  From the so called "the cosmological constant problem"$^{\scriptsize\cite{wei89}}$,  we know  that $\rho_{s}/\rho_{_0}\sim 10^{120}$.  Consequently,  we have $0<a_{s}\le 10^{-40}$, otherwise the energy density of our universe would be larger than the zero-point energy.
Therefore, our universe could be created from a limited region of $R_{s}\le 10^{-40} R_{_0}$ with an initial energy density $\rho_{s}$ at $z_s\le 10^{40}$ by a phase transition, taking place in a static primordial vacuum consisting of two balanced scalar fields$^{\scriptsize\cite{li11}}$. Such a scenario for the primordial universe may also explain  the observed lack of CMB power on the largest scales$^{\scriptsize\cite{liu11,liu13}}$.

   In $\mathnormal\Lambda$CDM,  the universe was created from a primordial singular point by a Big Bang. No coherent physics has been proposed for the Big Bang and  the following  Grand Unified Epoch  with $kT\sim 10^{19}$\,GeV. After  the  very hot earliest epoch,  the gravity, strong, weak and electromagnetic interactions were  created  in turn by a sequence of symmetry breakings following cooling temperatures$^{\scriptsize\cite{lin90,har00}}$.}  Contrarily, for a  DEMC  universe originated  from a cold primordial vacuum,  no Grand Unified Epoch exists. With the temperature increase caused by heating processes of the inflation,  the  breaks of symmetry  occurred in reverse order, i.e., the gravity interaction with inertial mass was created last at the end of the inflation. Through inflation, a considerable part of the primordial vacuum energy $E_{tot}$ was transformed into mechanical energy $E_{mech}$, as sketched by Figure\,\ref{fig:Etotal}. Except for temporary phase transitions, the universe after inflation with an inertial mass $M_{rest}$ has been and will be  expanding steadily with a constant rate $\sqrt{\epsilon}$.  The condensed universe always has a limited dimension and energy density, which can be readily described by current physics.\\
  \begin{figure}[t]
   \begin{center}
   \includegraphics[width=120mm,height=100mm,scale=0.6,angle=0]{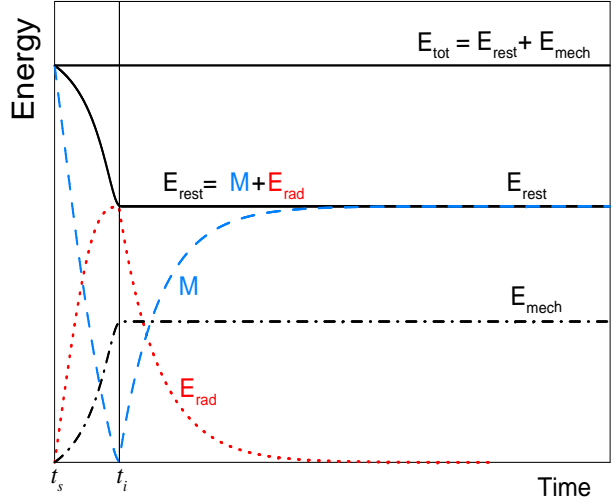}
   \vspace{-1.6cm} \caption{ \label{fig:Etotal} Conversion and conservation of energy in inflation and expansion eras.
   $t_s$: start time of inflation, $t_i$: end time of inflation.}
   \end{center}
\end{figure}

\noindent{\large\bf Discussion}

\noindent  An understandable  obstacle for the  proposed DEMC model to be accepted lies in its reliance on  not only  Einstein relativity,  but also Galilean relativity.  However,   that laws of physics must satisfy  Lorentz invariance is a misunderstanding.
In fact, non-relativistic  field theory is indeed essential for the many-body quantum theory in condensed matter physics and nuclear physics. The  Green-function theory developed for non-relativistic fields is an important constitution of the many-body quantum theory. Studying a certain collective object usually requires a privileged frame of reference.
For example, in solid state physics, the concerned  macroscopic object should always be at rest to the used coordinates,  and  to describe vibrations of lattices and lattice waves (phonons) of a crystal the canonical coordinates have to be used.  For collective phenomena of an object with respect to a specified frame of reference,  the condition of symmetry  is not  necessary the same as   in particle mechanics to the description of the motion of a single particle.
Although GR theory has to be applied to describe the motion of a particle in curved spacetime,  thermodynamics of an object in curved spacetime can be the same as classical  thermodynamics  except for the inclusion of rest mass together with all forms of relativistic mass-energy$^{\scriptsize\cite{mis73}}$.  In thermodynamics one confronts  two kinds of symmetries: Lorentz invariance for  locally determined parameters and Galilean invariance for collective properties.

There is a privileged frame of reference here for a homogeneous universe: the comoving  system with the Robertson-Walker metric. The cosmological expansion is the expansion of the comoving system itself, which is a kind of collective motion. But GR is a local theory valid for the dynamics of a bounded gravitational system constituted by massive particles, e.g. peculiar  motion of  a galaxy or merging of a  black hole binary,  not for scales lager than the relativistic causality restricted  by the limited speed of light.  Like Galilean invariance to  a local system,  Lorentz invariance to cosmic dynamics is also unsuitable.

A main reason to push Einstein to built the GR theory  is the difficulty in finding a space free of gravitational field for defining  an  inertial frame of reference$^{\scriptsize\cite{ein16}}$. Such a difficulty is not a trouble  any more for  the primordial  vacuum of a DEMC universe constituted by compensated attractive and repulsive fields with $\rho_{_g}=\rho_{_m}-\rho_{_\lambda}=0$, which  is just a privileged globally inertial and static frame.
The inflation, occurred in a limited region of the primordial  vacuum by a phase transition, while the region firstly transformed to be repulsive dominated and then
 the  repulsive field was converted back into attractive field again,  resulting in a part of the total energy   $E_{tot}$ of the transition region of the static vacuum into the mechanical energy $E_{mech}$ of the expanding universe.
Consequently,    Eq.\,(\ref{eq:ee})  representing mechanical energy conservation, $E_{mech}=E_k+E_p=$const,  can  be  defined
 in this initial static frame.

 In the initial static frame,  the primordial vacuum is homogeneous with maximum symmetry, and, in comoving coordinates,  the expanding universe  is also homogeneous with maximum symmetry.
The homogeneity of the universe enables us  to construct  a three-dimensional spacelike hypersurface, on which the expansion rate and acceleration are the same everywhere  at a given moment of time (\cite{mis73}, \S27.3). Therefore,  we can spell out the potential function of a universe at a given moment in Galilean framework, where relativistic retarded potentials cannot be used at all.
 Compared with $\rho_{_m}$ and $\rho_{_\lambda}$, the mass density of a massive particle is extremely large.  Even the mean mass density of a macroscopic object or an astrophysics system  (including the condensed halo of dark matter connected with the system)  is still very large,  thus the gravitational mass can be the same as the inertial mass of matter and the GR theory  can be applied. However, irrelevant to local phenomena, the cosmological expansion  drove by global gravity with gravitational density $\rho_{_m}-\rho_{_\lambda}$, which is not identical to the inertia density $\rho_{_m}+\rho_{_\lambda}$, can no longer use the GR theory.

Confusing the cosmological expansion with local processes is a common defect of   conventional models of relativistic cosmology.  To study the expansion of the universe by using particle dynamics, the test mass as an indicator of the expansion has to stay rest with respect to the comoving coordinates with pressure $P=0$, and its rest energy density $\rho$ already includes all forms of relativistic mass-energy.
The  first  Friedmann equation
\begin{equation}\label{eq:fe1}
\dot{a}^2=\frac{8\pi G}{3}\rho a^2+k \end{equation}
is deduced from the GR field equations and shows that the expansion of the universe is  governed  by the energy density $\rho$ alone. Whereas the second Friedmann equation
\begin{equation}\label{eq:fe2} \ddot{a}=-\frac{4\pi G}{3}(\rho+3P) a \end{equation}
is also deduced from  the GR field equations but contradictory to Eq.\,(\ref{eq:fe1}) in that the pressure $P$  can also drive the expansion directly.  However, contribution from the pressure  (including the radiation pressure at radiation dominated era)  to the cosmological expansion is already included in $\rho$, the pressure $P$ cannot have any more effect for cosmic expansion$^{\scriptsize\cite{oha13}}$.

 In cosmology,  cosmic processes have to be separated from local ones. The observed CMB  map of the whole sky has a homogeneous temperature $\bar{T}_{0}=2.725$\,K with
 small scale spots of temperature deviations  $\Delta T\sim\pm 10^{-5}\bar{T}_{0}$.
 The map of current cosmic thermal gravitational radiation of  dark sector should be completely homogeneous in the whole sky with the temperature $\bar{T}_0$.  Anisotropies of CMB, e.g. hot or cold spots with  angular scales $\sim 1\degree$ (the  horizon size),  are formed by local photons.
 One may think that cosmological scales were mixed with local ones for the tiny universe in early time. In fact, the early universe after a super-luminal inflation was already much larger than the particle horizon, and even more so than any bounded gravitational system. It is thus also necessary to separate global and local processes for early universe as well.
 Many contents in cosmology textbooks for early universe, e.g.   baryonsynthesis,  leptonsynthesis, nucleosynthesis, recombination, pertubation and structure formation, etc., are concerning just local astrophysics processes, where the GR theory can be applied.  Different from astrophysics, cosmology has a particular object,  the uniform and isotropic universe with a singer temperature $\bar{T}$ and density $\rho$ and investigates haw  $\bar{T}$ and $\rho$ vary with time,  ignoring their local fluctuations.
 To describe the cosmological expansion, cosmic dynamics, like thermal dynamics,  condensed matter physics and other collective physics, should be formulated in classical spacetime. However, it does not mean that the Einstein relativity is excluded in cosmology, because  its effects are already embodied in values of relevant  parameters, e.g. energy density $\rho$, redshift $z$,  etc.

There are two types of equally important gravitation in the universe: local gravity interaction in matter dominated systems and non-thermal GW,  both propagating in empty space at the speed of light$^{\scriptsize\cite{abb16,tan13}}$, and global gravitation of compensated dark energy and matter  with blackbody  gravitational phonons in Galilean spacetime.
The EM interaction and gravity interaction between massive particles were  successively  created during  inflation. At the end of inflation,  a small amount of the primordial vacuum energies were transformed into  massive particles, while most of them  into inertial dark matter and dark energy.  Macroscopic bodies and astronomical objects composed by massive and charged particles
are electrically neutral but attractive gravitation (matter)  dominated, whereas on cosmological scales the universe is gravitationally  neutral except in temporal phase transitions.
In the universe after inflation,  the non-thermal  EW  comes from accelerating charges and short-range interactions,  the thermal EW mainly from electrically neutral objects;
where the non-thermal GW comes from accelerating massive objects, and the thermal GW from the gravitational neutral cosmos.

   It is well known in cosmology that  the cosmic expansion  (expansion of  spacetime) can change the wavelength of light (cosmological redshift), but cannot let stars and galaxies themselves expand. Analogously, GW (ripples in spacetime) can change the motion of a photon, but also cannot change the dimension of a material body. In fact,   Gertsenshtein and Pustovoit$^{\scriptsize\cite{ger62}}$  in their paper proposing  the optical  interferometric detector of GW in 1962 pointed out that, it is not the non-relativistic body, but the light that sense the field of GW.  The detected relative phase change comes from changes of the {\sl optical} lengths of the interferometric arm lengths sensed by the light rays.  Because the light has to travel along the light cone, its spacetime interval is always equal to zero. The metric of a weak gravitational field can be expressed as $g_{\mu\nu}=\eta_{\mu\nu}+h_{\mu\nu}$,  where $\eta_{\mu\nu}$ is the flat Minkowski metric and $h_{\mu\nu}\ll 1$.
   For a weak plane GW  with the plus-polarization$^{\scriptsize\cite{che05}}$  traveling in the $z$ direction, the interval $\mbox{d}s$  of the light ray in the $x$ direction obeys
   $\mbox{d}s^2=-c^2\mbox{d}t^2+(1+h_{11})\mbox{d}x^2=0$, i.e.
   \[ c\,\mbox{d}t=\pm\sqrt{1+h_{11}}\,\mbox{d}x\,,\]
    indicating that  the light traveling is modulated by the space curved by GW.
 The LIGO experiment thus not only proves that GW can be generated by mergering of massive objects and travel in vacuum,  but also reveals profound connections between  electromagnetic interaction and gravitation by the process  that  laser  photons traveling in empty space  are guided by GW. If the detection of EW by Hertz demonstrated the uniformity of electromagnetic interaction and thermal radiation, the detection of GW further indicates the uniformity of  electromagnetic interaction, gravitation and thermal radiation.
 For retaining the universe's  global thermal equilibrium  and energy conservation,  the cosmological gravitational radiation and the connection between local EW and global GW have to be
 taken into account.

In searching for  gravitation theory, Einstein used the energy-momentum conservation law as a constraint on the theory$^{\scriptsize\cite{pai82}}$.
The field equation with the cosmological constant $\mathnormal\Lambda$ proposed by Einstein$^{\scriptsize\cite{ein17}}$  on relativistic cosmology relied upon, except for covariance, the following aspects: the universe being spatially finite and static; the cosmic constant $\mathnormal\Lambda$ representing a mean matter field; and the total mass of universe as a finite constant.
It has to be pointed out that only for a static universe the above aspects can be compatible with each other.  Einstein is right to give up the cosmological constant when the expansion of the universe was observed, otherwise, as in $\mathnormal\Lambda$CDM with a density of dark energy  $\rho_{_\lambda}=\mathnormal\Lambda/8\pi G=$ const, the total energy of the universe is infinitely increasing with time, $E_{rest}=\rho V\rightarrow \infty$, severely violate  energy conservation.
One may argue that the GR field equations with a  constant term satisfy the covariance requirement, and, then,  the energy conservation from the Noether theorem.   However,  the first necessary   precondition of the conservation law and Noether theorem is the concerned system must be a closed one, an expanding system with a constant density of energy cannot be closed at all.
To built a rational physics model for a closed system, such as for the universe,  the conservation law is a stronger constraint  than the condition of covariance,  or, in other words, physical conditions  in physics take higher  priority over  mathematical ones.

In DEMC scenario, we take  dark energy as another fundamental  composition of the universe besides matter.  Different from $\mathnormal\Lambda$CDM, the energy density $\rho_{_\lambda}$ is not a constant,  the rest energy $E_{rest}$,  and the mechanical energy $E_{mech}$ of the expanding universe are conserved separately.
The repulsive dark energy  should be a continuous medium, not composed of massive particles. To keep the universe in perfect  thermal equilibrium, dark matter should also be a  continuous medium tightly coupled with dark energy, where  in cosmic dynamics  the small amount of mass of discrete massive particles  has to be smoothed into a homogenous continuum.
   The assumption of DEMC cosmology, namely the universe being constituted  by two opposite signed and compensated continuous gravitational fields,  provides a simple and sound physical foundation for cosmology with homogeneity, isotropy, energy conservation, thermal equilibrium, non-locality, and non-singularity.




\end{document}